\documentclass[seceq]{ptptex}

\usepackage{graphicx}

\title{
Hard QCD Probes to Quark-Gluon Plasma%
}

\author{
Carlos A. \textsc{Salgado}\footnote{ e-mail address: carlos.salgado@cern.ch}  
}

\inst{
Dipartimento di Fisica, Universit\`a di Roma ``La Sapienza"
and INFN, Roma, Italy\\
and\\
Departamento de F\'\i sica de Part\'\i culas and IGFAE,
Universidade de Santiago de Compostela,  Santiago de Compostela, Spain
}

\abst{
Completely unexplored regimes of QCD, dominated by high-density/temperature effects, are available in heavy ion experiments at collider energies. The successful RHIC program shows how relevant the high transverse momentum part of the spectrum is for the characterization of the properties of the created medium. It points, as well, to interesting properties of the nuclear wave function at small fraction of momentum $x$, probably dominated by saturated color fields. In both domains, the imminent LHC program will provide a phase space enlarged by orders of magnitude with respect to those studied at RHIC. I will review the present status of hard probes in heavy ion collisions as well as the expectations for the LHC.}

\begin{document}

\maketitle

\section{Introduction}

High-energy heavy-ion collisions are the experimental tools available to study how collective behavior appears in QCD. The spacial extension of the nuclei ensures that high density states are produced in an {\it extended} region of space, which is a prerequisite for collectivity to manifest. The lifetime of the formed medium is rather short -- of the order of its transverse size -- and different signals to characterize its properties are proposed. From a theoretical point of view, a good signal is that in which good control on the medium property as extracted from the modification of a well calibrated process is possible. Although no perfect signal exists hard probes are at collider energies in an excellent position to make the bridge between the experimental data and the theoretical characterization of the medium properties.

The first example of a hard probe was proposed in the 80's by Matsui and Satz \cite{Matsui:1986dk}: the screened (non-confining) potential of a charm-anticharm pair in a thermal medium makes the hadronization process into a $J/\Psi$ very unlikely and a suppression of the $J/\Psi$ yield is expected. The $J/\Psi$ suppression, as well as the suppression of other charmonia states has been observed experimentally \cite{jpsiexp}. 

\section{Hard processes in hadronic collisions}

A typical hard cross section can be written in the factorized form
\begin{equation}
\sigma^{AB\to h}=
f_A(x_1,Q^2)\otimes f_B(x_2,Q^2)\otimes \sigma(x_1,x_2,Q^2)\otimes D_{i\to h}
(z,Q^2)\, ,
\label{eqhard}
\end{equation}
where the short-distance perturbative cross section, $\sigma(x_1,x_2,Q^2)$, is computable in powers of $\alpha_s(Q^2)$ and the long-distance terms are non-perturbative quantities involving scales ${\cal O}(\Lambda_{\rm QCD})$ but whose evolution in $Q^2$ can be computed perturbatively. More specifically, the proton/nuclear parton distribution functions (PDF), $f_A(x,Q^2)$, encode the partonic structure of the colliding objects at a given fraction of momentum $x$ and virtuality $Q$; and the fragmentation functions (FF), $D(z,Q^2)$, describe the hadronization of the parton $i$ into a final hadron $h$ with a fraction of momentum $z$. In the nuclear case, these are the quantities which are modified when the {\it extension} of the colliding system interferes with the dynamics, while the short-distance part is expected to remain unchanged if the virtuality is large enough. These modifications could involve the non-perturbative initial condition as well as the evolution equations. In this last case, non-linear terms become important.

A conceptually simple example is the above mentioned case of the $J/\Psi$, whose production cross section can be written as
\begin{equation}
\sigma^{hh\to J/\Psi}=
 f_i(x_1,Q^2)\otimes f_j(x_2,Q^2)\otimes
\sigma^{ij\to [c\bar c]}(x_1,x_2,Q^2)
 \langle {\cal O}([c\bar c]\to J/\Psi)\rangle\, ,
\end{equation}
where now $ \langle {\cal O}([c\bar c]\to J/\Psi)\rangle$ describes the hadronization of a $c\bar c$ pair in a given state (for example a color octet) into a final $J/\Psi$. This is a purely non-perturbative quantity, which, as it was said before,  is expected to vanish when the medium is hot  \cite{Matsui:1986dk}. This modification, being non-perturbative, lacks of good theoretical control, making difficult the interpretation of the experimental data.

From the computational point of view, a theoretically simpler case is the modification of the {\it evolution} of both the parton distribution and the fragmentation functions in a dense or finite--temperature medium. This needs of large scales $Q^2$ (small-$x$) to access the slow logarithmic dependences involved.

\begin{figure}
\begin{minipage}{0.55\textwidth}
\begin{center}
\includegraphics[width=\textwidth]{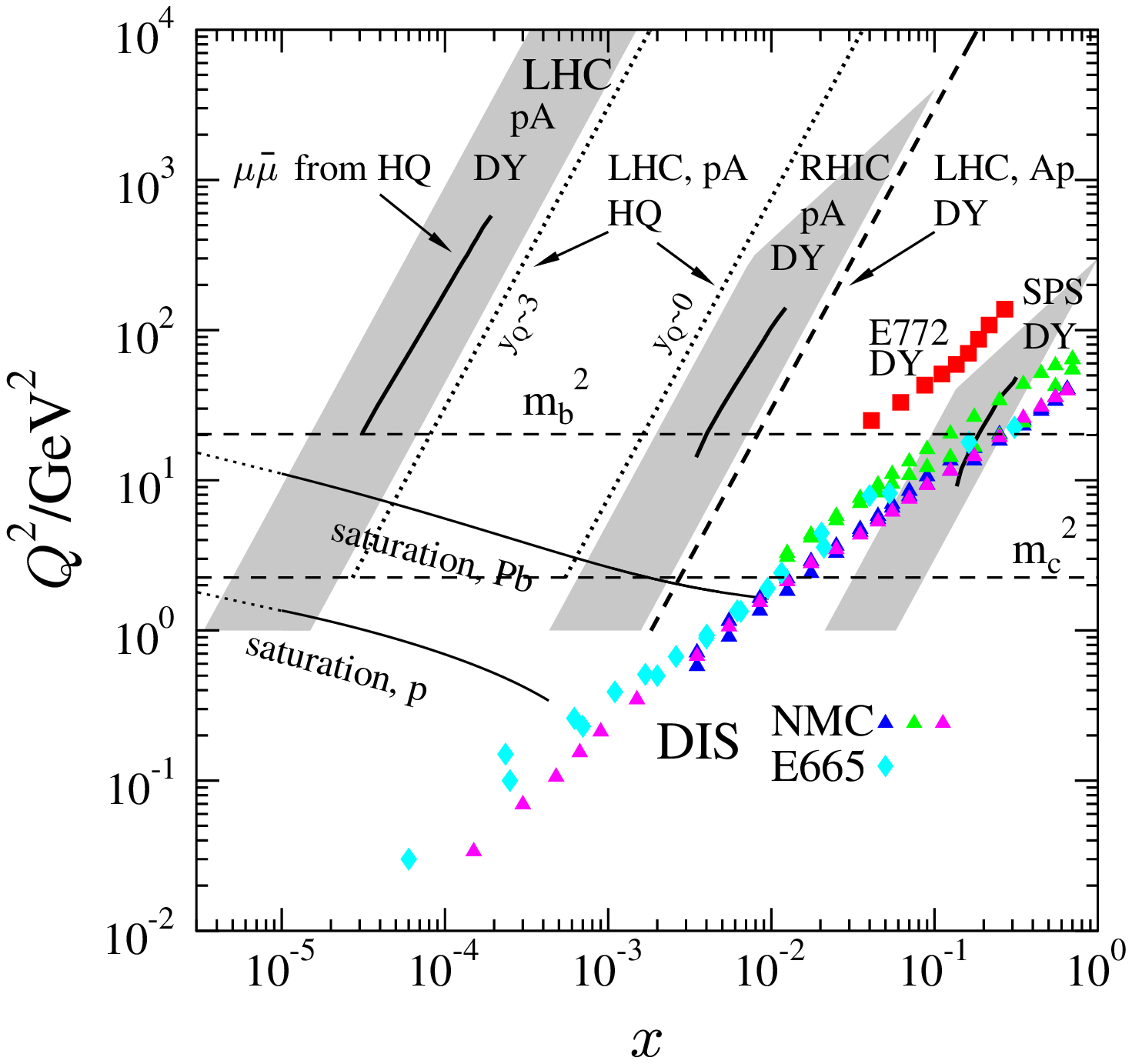}
\end{center}
\end{minipage}
\hfill
\begin{minipage}{0.45\textwidth}
\begin{center}
\includegraphics[width=\textwidth]{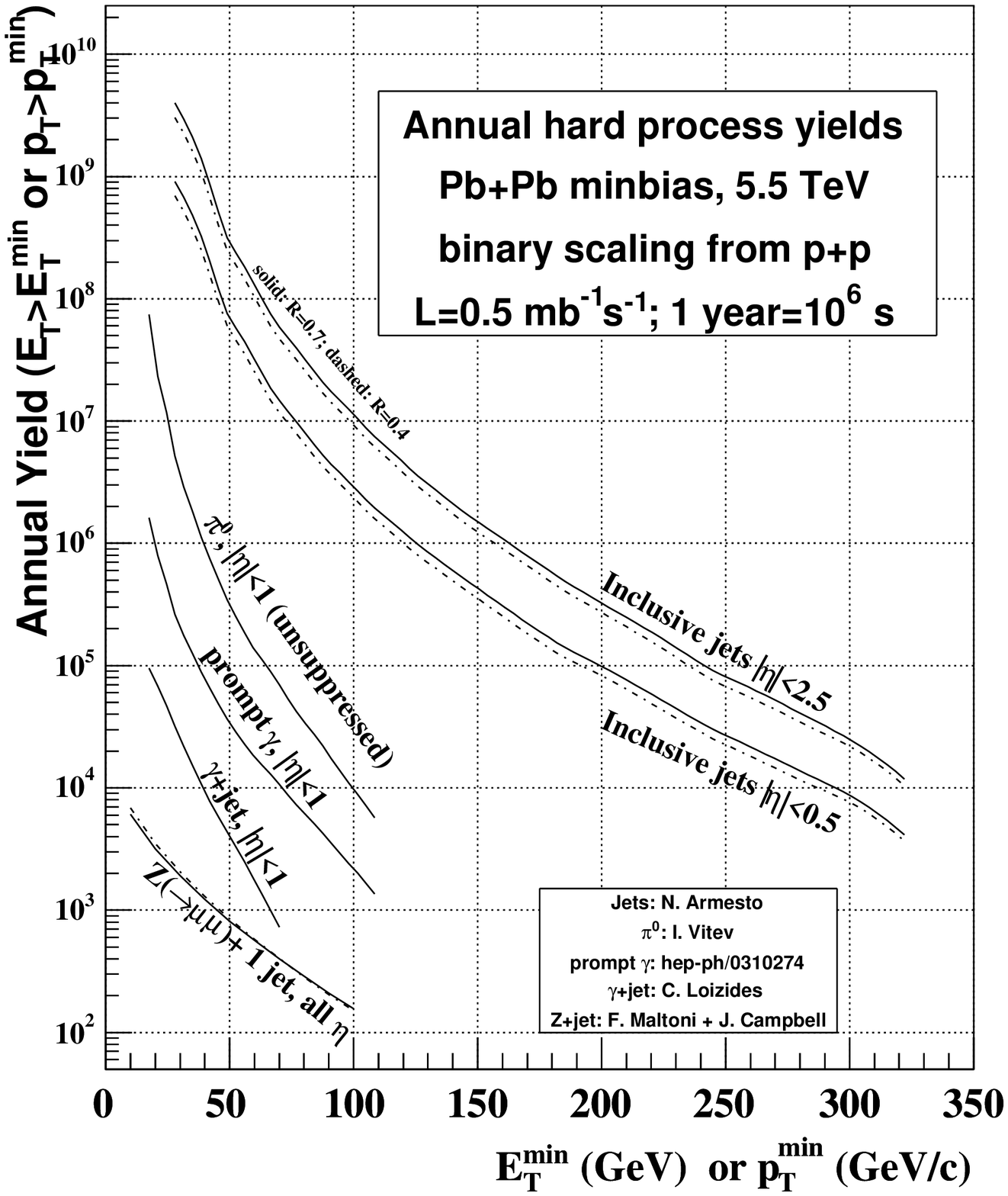}
\end{center}
\end{minipage}
\vskip -1cm
\caption{Left: $(x,Q^2)$ kinematic plane reachable at the LHC -- figure from \protect\cite{Accardi:2004be}. Right: Integrated yields of different high-$p_t$ processes at the LHC -- figure from \protect\cite{Jacobs:2005tq}.}
\label{fig:kinem}
\end{figure}

In Fig. \ref{fig:kinem} the kinematic regimes reachable at the LHC both in $x$ and transverse momentum are presented. While RHIC kinematics allowed, for the first time, to do real hard probe studies in heavy ion collisions with well calibrated processes, the most important step forward at the LHC is the sensitivity to in-medium modifications of the QCD evolution thanks to an enhanced kinematical reach of three orders of magnitude in $x$ and more than one in transverse momentum.

\section{Nuclear parton distribution functions}

Essential in any calculation of hard processes is a good knowledge of the PDFs. The usual way of obtaining these distributions is by a global fit of data on different hard processes 
(mainly deep inelastic scattering, DIS) to obtain a set of parameters for the initial, non-perturbative, input $f(x,Q^2_0)$ to be evolved by DGLAP equations \cite{DGLAP}.

In the nuclear case, the initial condition, $f_A(x,Q^2_0)$, is modified compared to the proton. Moreover, at small enough $x$, non-linear corrections to the evolution equations are expected to become relevant. Global DGLAP analyses, paralleling those for free protons are available \cite{Eskola:1998iy,Eskola:1998df,Hirai:2001np,Hirai:2004wq,deFlorian:2003qf,Eskola:2007my}. These studies fit the available data on DIS and Drell-Yan with nuclei providing the needed benchmark for additional mechanisms.

\begin{figure}[tbh]
\centering
\centering\includegraphics[width=0.6\textwidth]{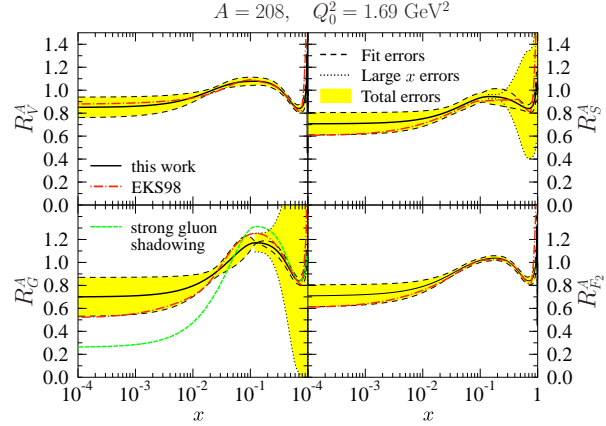}
\caption{Ratios of nuclear to free proton PDFs for different flavors at the initial scale $Q_0^2$=1.69 GeV$^2$ from \protect\cite{Eskola:2007my} with error estimates. The green line in the gluon panel is an attempt to check the strongest gluon shadowing supported by present data. }
\label{fig:nPDF}
\end{figure}

 The most recent \cite{Eskola:2007my} of the DGLAP analyses of nuclear PDFs is shown in Fig. \ref{fig:nPDF}, including the corresponding error estimates. An important issue, partially visible in Fig. \ref{fig:nPDF}, is that present nuclear DIS and DY data can only constrain the distributions for $x\gtrsim 0.01$ in the perturbative  region. By chance, this region covers most of the RHIC kinematics, so that, the description of e.g. $J/\Psi$-suppression or inclusive particle production in  dAu collisions as given by the nuclear PDFs can be taken as a check of universality of these distributions. These checks present a quite reasonable agreement with data \cite{Vogt:2004hf}, but some extra suppression for the inclusive yields at forward rapidities is probably present. The strong gluon shadowing plotted in Fig. \ref{fig:nPDF} improves the situation at forward rapidities without worsening the fit of DIS or DY data -- $\chi^2/{\rm dof}<1$. Whether a DGLAP analysis can accommodate all sets of data is an open question, but the finding in Ref. \cite{Eskola:2007my} are encouraging. A suppression at forward rapidities was also predicted in terms of saturation of partonic densities \cite{satur}. 
 
\subsection{Saturation of partonic densities: the CGC}

When the partonic densities are large enough (at small-$x$ and/or large-$A$) non-linear terms in the evolution equations become sizable, and eventually dominate. These terms are needed to tame the growth of these densities which, otherwise, will lead to a violation of the S-matrix unitarity.  Although several early attempts to compute these non-linearities exist \cite{Gribov:1984tu,Mueller:1985wy} the most developed formalism is, nowadays, based on a semiclassical approach \cite{McLerran:1993ni}, in which the nuclear wave function at large energies is treated as an ensemble of classical color field configurations. The evolution with energy of these configurations is known -- the B-JIMWLK equations \cite{jimwlk}, whose mean field limit acquires a simple form \cite{Kovchegov:1999yj}. In the dilute regime, where the non-linear terms are negligible, the BFKL equation is recovered. This approach is called the Color Glass Condensate.

\begin{figure}
\begin{minipage}{0.35\textwidth}
\begin{center}
\includegraphics[width=\textwidth]{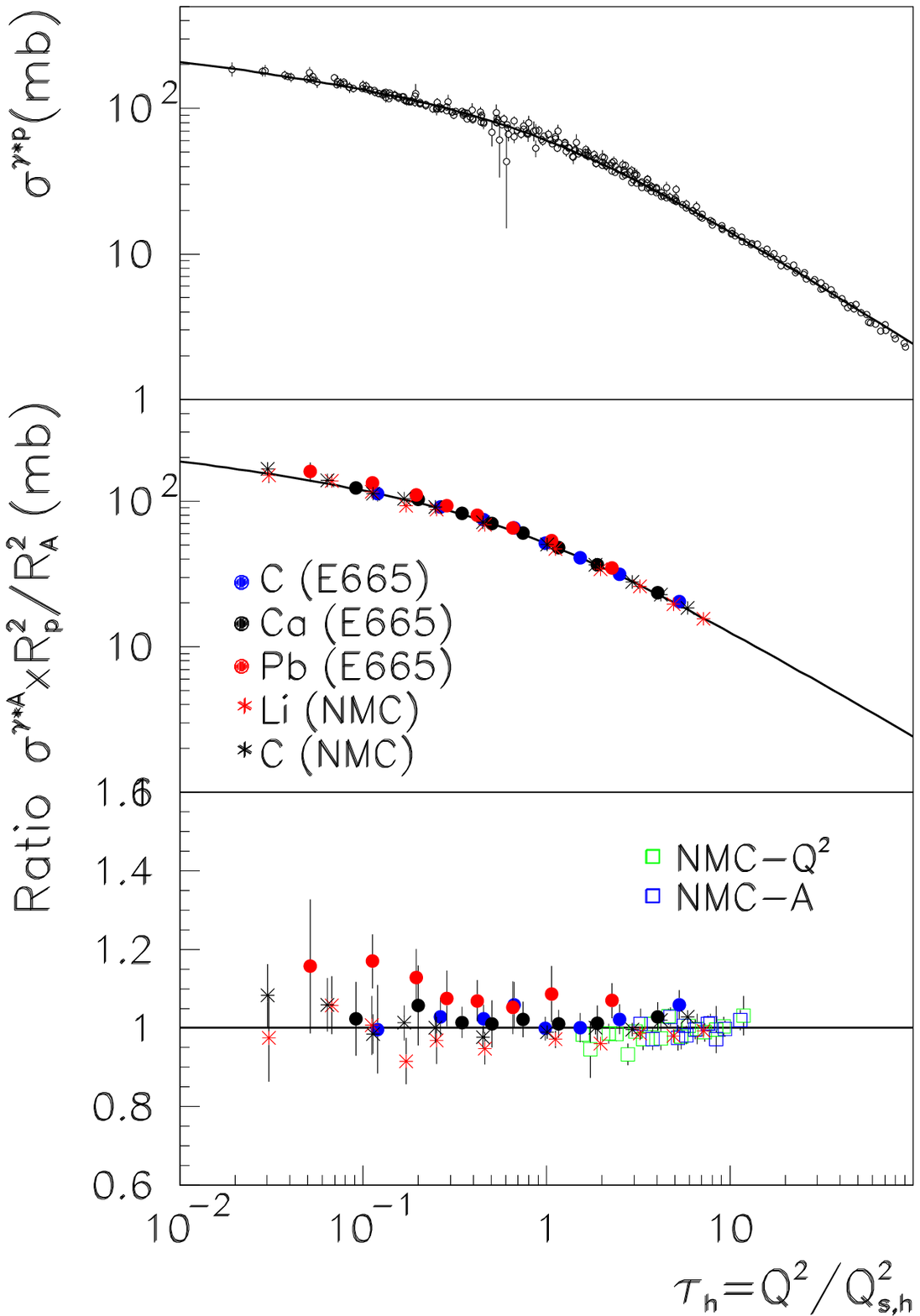}
\end{center}
\end{minipage}
\hfill
\begin{minipage}{0.65\textwidth}
\begin{center}
\includegraphics[width=0.8\textwidth]{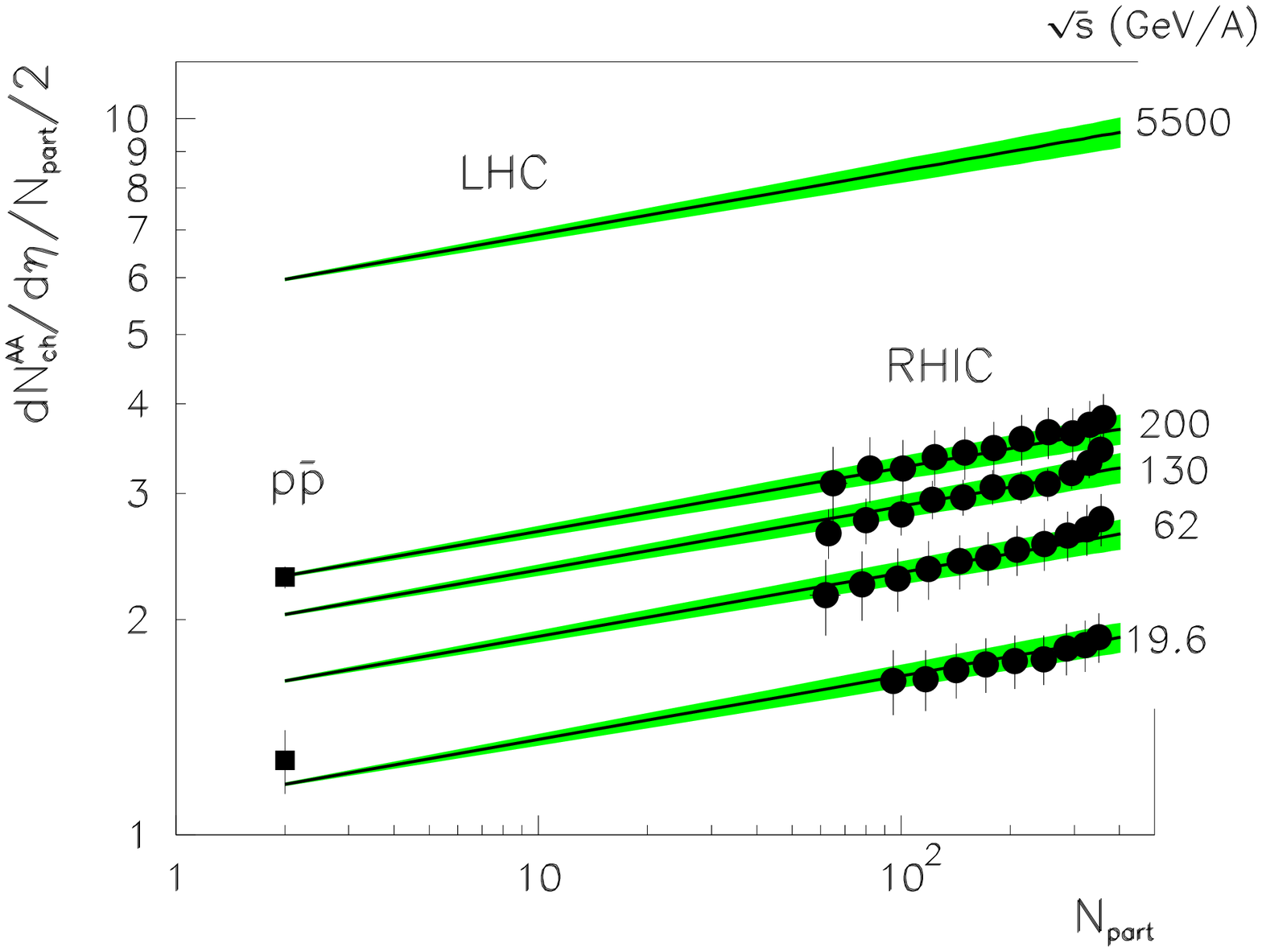}
\end{center}
\end{minipage}
\caption{Left: Geometric scaling in lepton-proton \protect\cite{Stasto:2000er} and lepton-nucleus \protect\cite{Armesto:2004ud} data. Right: Central rapidity multiplicities in $p\bar p$ and AuAu collisions  \protect\cite{Back:2004je} and the description by geometric scaling \protect\cite{Armesto:2004ud}.}
\label{fig:geoscal}
\end{figure}

Although a description of the experimental data by directly solve and fit the non-linear evolution equations is still missing -- see, however, Ref. \cite{Albacete:2007hg} -- much of the phenomenological work has pursued the existence of known properties of the asymptotic solution of the equations. In particular, the CGC predicts a {\it geometric scaling} in which the partonic distributions are only a function of the ratio $Q^2/Q^2_{\rm sat,A}(x)$, where the saturation scale $Q_{\rm sat,A}(x)$ contains all the $x$- and $A$-dependences. This feature is compatible with experimental data on proton-\cite{Stasto:2000er} and nuclear-DIS \cite{Armesto:2004ud} and could explain the multiplicities \cite{Kharzeev:2000ph,Armesto:2004ud} measured in hadronic collisions -- see Fig.  \ref{fig:geoscal}.

\section{High-$p_t$ studies in heavy ion collisions: Jet quenching}

Quarks or gluons produced at high transverse momentum in elementary collisions
build up their hadronic wave function by emitting partons, mostly gluons, 
reducing their large virtuality ($Q\sim p_t$) to a
typical hadronic scale. These emitted particles form QCD parton showers which
can be experimentally observed as jets. The
properties of these objects can be computed by resumming the relevant logarithms
originated by the large difference in the scales of the problem. One example of these resummations are the DGLAP evolution equations \cite{DGLAP}
describing the virtuality dependence of the fragmentation functions.

\begin{figure}
\begin{minipage}{0.5\textwidth}
\begin{center}
\includegraphics[width=0.75\textwidth,angle=-90]{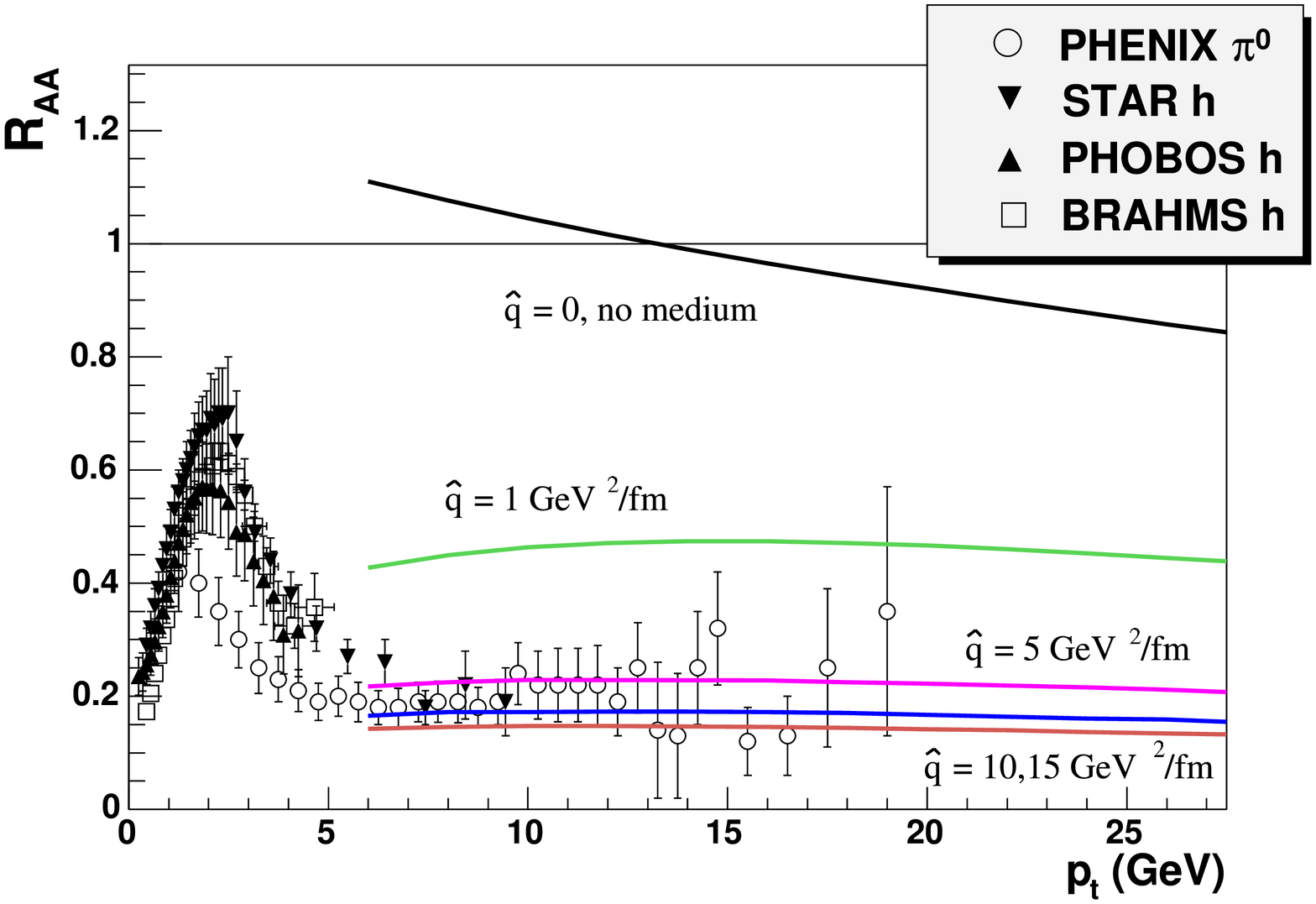}\\
\vskip 0.2cm
\centerline{  }
\end{center}
\end{minipage}
\hfill
\begin{minipage}{0.5\textwidth}
\begin{center}
\includegraphics[width=0.8\textwidth]{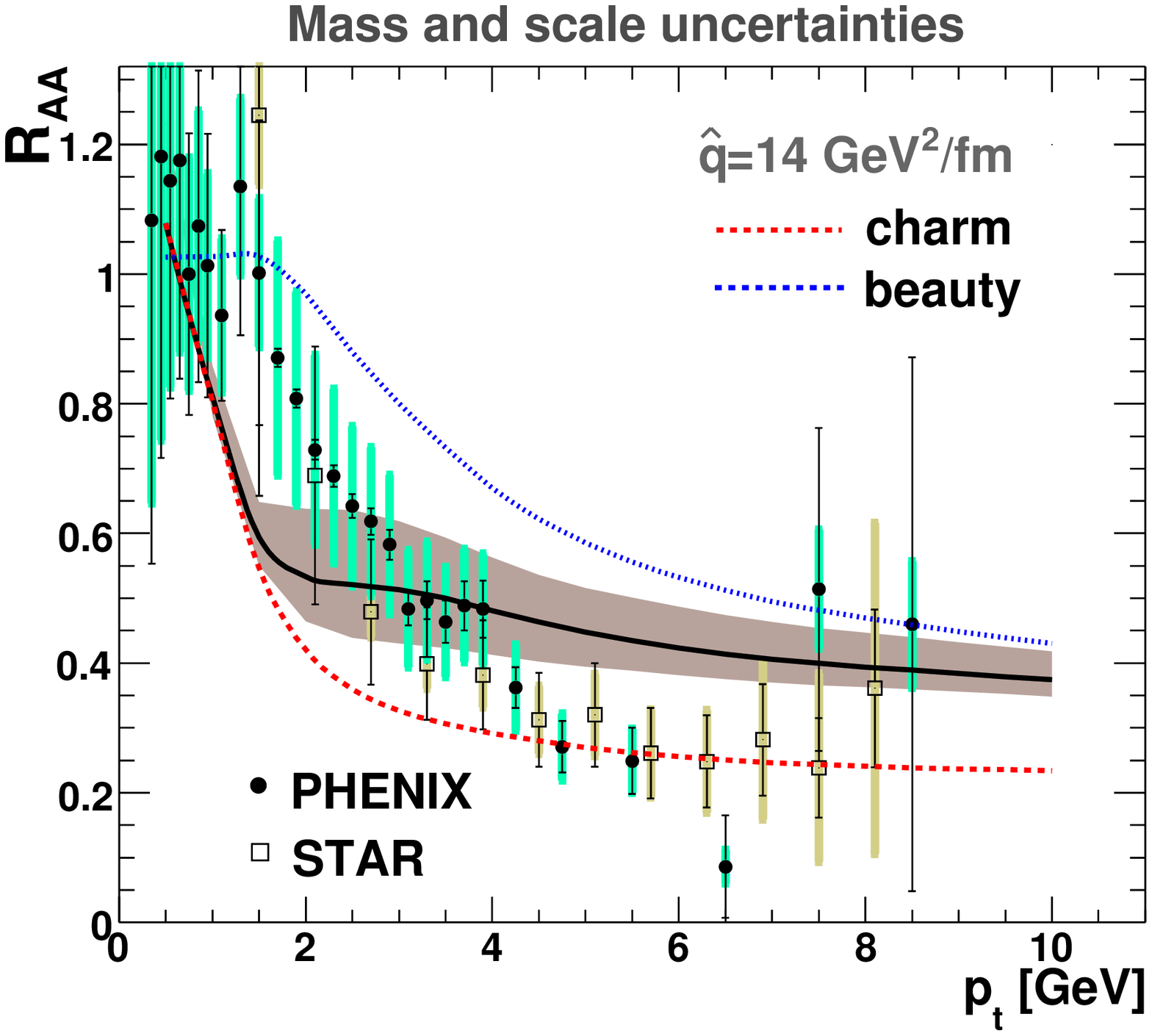}
\end{center}
\end{minipage}
\caption{Left: Nuclear modification factor, $R_{AA}$, for light hadrons in central
AuAu collisions \protect\cite{Eskola:2004cr}. Data from \protect\cite{Adcox:2001jp}. Right: $R_{AA}$ for non-photonic electrons with the corresponding uncertainty from the perturbative benchmark on the relative $b/c$ contribution \protect\cite{Armesto:2005mz}. Data from \cite{Abelev:2006db,Adare:2006hc}}
\label{fig:supp}
\end{figure}

Jet structures are expected to be modified when the evolution takes place into a thermal medium. The associated effects are generically known under the name of {\it jet quenching} and the dominant mechanism is the medium-induced gluon radiation  \cite{Baier:1996sk,Zakharov:1997uu,Wiedemann:2000za,Gyulassy:2000er,Wang:2001if}. This induced radiation modifies the vacuum splitting functions producing additional energy loss and broadening of the jet transverse profile. The simplest observational prediction from this formalism is the suppression of the inclusive particle production at high-$p_t$. This suppression can be traced back to a medium-modification of the fragmentation function $D_{i\to h}(z,Q^2)$ at relatively large values of $z\gtrsim 0.5$  -- the most relevant ones in Eq. (\ref{eqhard}) due  to the bias effect induced by the steeply falling perturbative spectrum \cite{Baier:2001yt}. Most of the present phenomenology assumes a medium modification of the fragmentation function due to energy loss
\begin{equation}
D_{i\to h}^{\rm med}(z,Q^2)=P_E(\epsilon)\otimes D_{i\to h}(z,Q^2)
\label{eqff}
\end{equation}
neglecting any modification of the virtuality dependence of the vacuum fragmentation function \cite{Wang:1996yh}, and where  $P_E(\epsilon)$ is computed in the independent  gluon emission approximation\cite{Baier:2001yt,Salgado:2003gb}. 
The medium-induced energy loss probability distribution $P_E(\epsilon)$ -- known as {\it quenching weights}, QW -- depends only on the in-medium path-length of the hard parton and the transport coefficient $\hat q$. The length is given by geometry and it is not a free parameter of the calculation -- although different geometries, including expansion, hydrodynamics, etc. could lead to slightly different results \cite{Renk:2006pk}. The transport coefficient encodes all the properties of the medium accessible by this probe and can be related to the average transverse momentum gained by the gluon per mean free path in the medium. Taking it as a free parameter of the calculation and fitting available data, a value of \cite{Eskola:2004cr,Dainese:2004te}
\begin{equation}
\hat q=5....15\, {\rm GeV}^2/{\rm fm} 
\label{eq:qhat}
\end{equation}
is obtained.
The quality of the fit can be seen in Fig. \ref{fig:supp}. The large uncertainty in the determination of $\hat q$ is a consequence of the large opacity of the medium, which together with the bias effect mentioned above, leads to a surface-dominated emission probability for the particles escaping the medium \cite{Eskola:2004cr,Dainese:2004te,Muller:2002fa}. The situation can be improved by measuring the identity dependence of the energy loss (e.g. with heavy quarks) and/or by detecting the structure of the associated induced radiation. Both type of measurements do not involve any new parameter. 

Heavy meson production is measured at RHIC through their decay into electrons. The strong suppression measured in central AuAu collisions \cite{Abelev:2006db,Adare:2006hc} contains a mixture of charm and beauty contributions not yet under good theoretical control. The description of the data within the formalism is reasonable \protect\cite{Armesto:2005mz} -- see Fig. \ref{fig:supp} -- but an experimental separation of both contribution will help to understand whether other effects \cite{Wicks:2007sn} are at work .

\subsection{Jets}

The most promising signal of the dynamics underlying jet quenching is the study of the modifications of the jet structures \cite{Salgado:2003rv} in which the characteristic angular dependence of the associated medium-induced radiation \cite{Baier:1996sk,Zakharov:1997uu,Wiedemann:2000za,Gyulassy:2000er,Wang:2001if, Polosa:2006hb} should be reflected. Experimentally, the main issue to overcome is the jet energy calibration in a high-multiplicity environment where small-$p_t$ cuts and more or less involved methods of background subtraction will be needed. From a theoretical point of view, identifying signals with small sensitivity to these subtractions is of primary importance \cite{Salgado:2003rv}. Due to these limitations, jet studies are not possible in AuAu collisions at RHIC but will be abundant at the LHC up to transverse energies of several hundred GeV -- see Fig. \ref{fig:kinem}. In the meantime, jet-like structures are being studied at RHIC by means of two- and three-particle correlations. 

\begin{figure}
\begin{minipage}{0.48\textwidth}
\begin{center}
\includegraphics[width=0.68\textwidth,angle=-90]{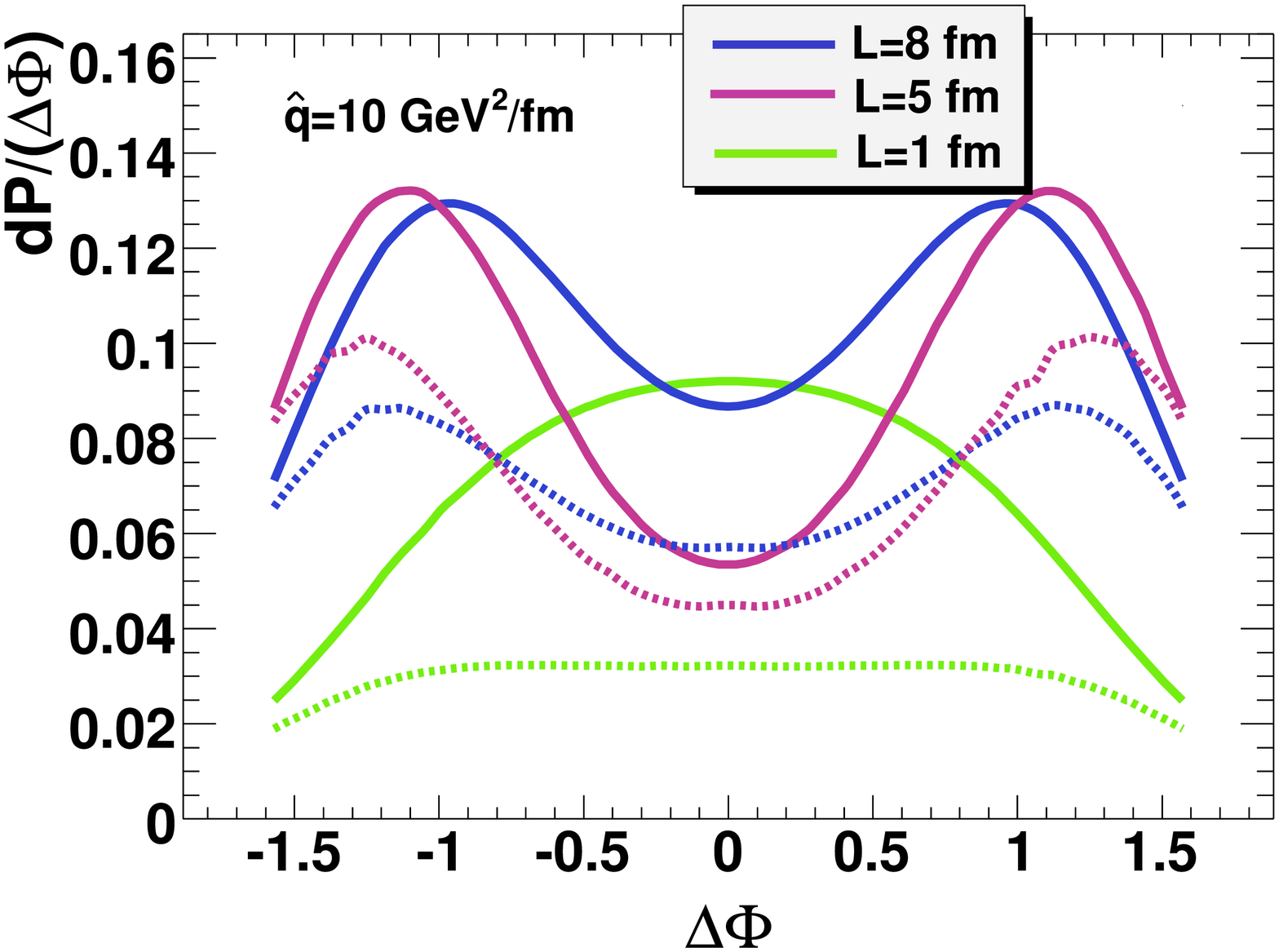}
\end{center}
\end{minipage}
\hfill
\begin{minipage}{0.48\textwidth}
\begin{center}
\includegraphics[width=0.68\textwidth,angle=-90]{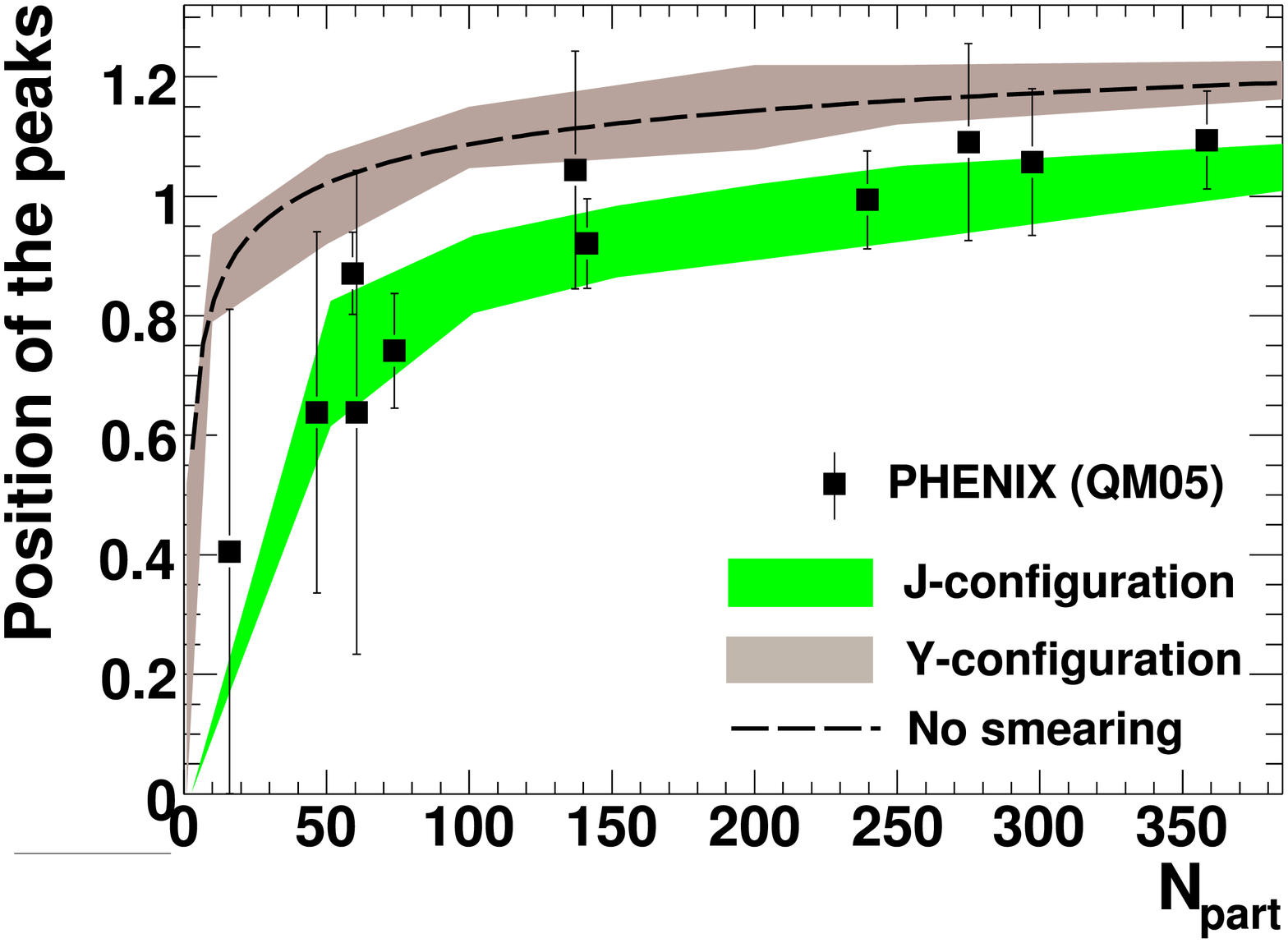}
\end{center}
\end{minipage}
\caption{Left: The probability of one splitting \cite{Polosa:2006hb} as a function of the azimuthal angle $\Delta\Phi$ for a gluon jet of $E_{\rm  jet}=7$ GeV. Right: Position of the peaks 
 and comparison with PHENIX data \cite{Adler:2005ee}.}
\label{fig:opaque}
\end{figure}

An important step forward is the first measurement of two particle azimuthal correlations at large transverse momentum, with negligible combinatorial background \cite{Adams:2006yt}. These data support the picture of a very opaque medium with large energy losses, but with a broadening of the associated soft radiation hidden underneath the cut-off. Lowering this transverse momentum cut-off needs of a good control on the background subtraction, but the different collaborations agree in the presence of non-trivial angular structures \cite{Adler:2005ee}: the two-particle-correlation signal around the direction opposite to the trigger particle presents a dip in central collisions, in striking contrast with the typical Gaussian-like shape in proton-proton or peripheral AuAu collisions. In the presence of an ordering variable (as virtuality or angular ordering in the vacuum parton shower) the implementation of the usual Sudakov form factors to the medium-induced gluon radiation produce similar angular structures for energies $\omega\lesssim 2\hat q^{1/3}\sim 3$ GeV for central AuAu \cite{Polosa:2006hb} -- see Fig. \ref{fig:opaque}. 

The medium-induced gluon radiation assumes that most of the energy is lost by radiation with negligible deposition in the medium. On the opposite limit, if a large fraction of the jet energy is deposited fast enough into a hydrodynamical medium it will be diffused by sound and/or dispersive modes. For very energetic particles, traveling faster than the speed of sound in the medium, a shock wave is produced with a characteristic angle which could be at the origin of the measured structures \cite{conical}. Another interpretation of this effect is in terms of Cherenkov radiation \cite{cherenkov}. 

Additional information comes from the {\it near side} two-particle correlations, where an elongation in the longitudinal direction of the jet signal around the trigger particle is observed \cite{Jacobs:2005pk}. Although not fully understood, these data points to a coupling between the in-medium jet evolution and the presence of hydrodynamical flow fields \cite{Armesto:2004pt}. The study of flow fields with jet measurements would became possible in this manner. 

Independently on the actual interpretation of these findings, the jet-like particle correlations at RHIC provide an experimental measurement on the amount of energy deposition in the medium and the parton shower evolution. Jet studies at the LHC are ideal tools to further unravel the underline dynamics of jet quenching in heavy-ion collisions and to study the medium properties with unprecedented precision.

\section*{Acknowledgements}
I would like to thank the organizers of the YKIS2006 on "New Frontiers on QCD" for the nice atmosphere during the workshop. This work is supported by the FP6  of the European Community under the contract MEIF-CT-2005-024624.

%


\end{document}